# New CuSO$_4$-related high-temperature polymorph of Ag$^{II}$SO$_4$

Mateusz Domański,*[a] Zoran Mazej,[b] and Wojciech Grochala[a]

This work is dedicated to 100$^{th}$ birthday anniversary of prof. R. Hoppe

**Abstract:** Silver(II) compounds exhibit powerful oxidizing properties and strong magnetic superexchange. AgSO$_4$ is a rare fluorine-free salt of Ag(II) which found some application in organic chemistry. Here, we report a discovery of a new AgSO$_4$ polymorph (β). The distinct nature of the two polytypes of AgSO$_4$ is established using powder x-ray diffraction, vibrational spectroscopy and theoretical calculations. The β polymorph crystallizes in the monoclinic system ($P2_1/n$) and shows structural similarities with CuSO$_4$. DFT calculations indicate very small differences in the energy of the two polymorphs AgSO$_4$, but the relative stability of the β polymorph should increase with temperature. The monoclinic distortion of the orthorhombic CuSO$_4$ prototype originates from an unprecedented strong antiferromagnetic interaction between Ag sites along the unit cell diagonal.

## Introduction

Silver(II) compounds have gained an increased attention over the last two decades, mostly due to their strongly oxidizing properties and interesting magnetic properties. Most of them comprise a fluoride ligand, as Ag(II) tends to disproportionate in oxide environment. For example, the silver oxide of AgO stoichiometry is a prototypical example of a Ag(I)/Ag(III) compound.[1] Ag$_3$O$_4$ is a mixed-valent Ag(II)/Ag(III) compound, while Ag$_2$O$_3$ contains only Ag(III).[2-6] In addition, Ag$_6$O$_2$ and Ag$_2$O are also known.[6] On the other hand, copper(II) oxides and more complex oxo salts such as sulphate or nitrate, have been known even in antiquity. Remarkably, until 2010 no fluorine-free silver(II) analogues were known. A question of a possible existence of Ag(II)SO$_4$ was raised based on analysis of thermodynamic data[7], as the earlier density functional theory (DFT) calculations suggested that molecular AgSO$_4$ in the gas phase would have spin mostly on a sulphate group[8]. Subsequently, in a theoretical study based on DFT non-spin-polarized calculations, Derzsi and coworkers predicted the AgSO$_4$ crystal structure based on seventeen plausible structural models; the lowest enthalpy structure contained genuine Ag(II)[9]. This prediction was confirmed via synthesis of silver(II) sulfate (labelled here as α-AgSO$_4$) and subsequently of its monohydrate derivative[10,11]. These studies encompassed synthetic methods, analysis of vibrational and EPR spectra, and magnetic susceptibility profiles. Subsequently, reactivity of AgSO$_4$ was explored, including its strong one-electron oxidizing capability[12] which is essential in further applications of this compound in organic chemistry.

With s=1/2 spin carried by Ag(II) centers, α-AgSO$_4$ is a quasi-1D antiferromagnet. The spin superexchange integral $J$ between the nearest neighbours for AgSO$_4$ was measured to be –9.5 meV per one Ag while the value calculated with a DFT method equals –6.2 meV (we note that the intra- and inter-chain superexchange integrals were in fact averaged in the mentioned calculations[13]). The superexchange *via* sulphate groups in α-AgSO$_4$ is predicted to bypass sulfur atom and involve an unusual direct O⋯O spin polarization (*cf.* Fig.1 from the work of Malinowski *et al.*[10]).

In 2012, the crystal structure of silver(II) sulphate proposed by Malinowski *et al.*[10] was redetermined[13]. α-AgSO$_4$ crystallizes in a $C2/c$ space group, and is related to a high-temperature polymorph of PdSO$_4$[13]. α-AgSO$_4$ does not undergo any pressure-induced structural phase transitions for p up to about 30 GPa[13]. Herein we describe the new polymorph of AgSO$_4$ labelled as β-AgSO$_4$, which was obtained using wet chemistry approach. We compare its structure and properties to those of the known α-AgSO$_4$ as well as to related CuSO$_4$[14]. The third member of the Cu, Ag, and Au triad (*i.e.*, AuSO$_4$) is also a true metal(II) sulphate. However, its crystal structure is unique in that it contains dumpbell shaped [Au$_2$]$^{4+}$ cations linked by [SO$_4$]$^{-2}$ anions.[15]

## Results and Discussion

### Crystal structure of β-AgSO$_4$

β-AgSO$_4$ is formed in an acid displacement reaction starting from the AgF$_2$ precursor, where weak acid HF is eliminated by anhydrous H$_2$SO$_4$ poured onto AgF$_2$ cooled in liquid-nitrogen:

AgF$_{2(s)}$ + H$_2$SO$_{4(l)}$ → AgSO$_{4(s)}$ + 2 HF$_{(g)}$↑

The key element in obtaining the new polymorph of AgSO$_4$, as opposed to the previously known form, is to place the reaction vessel containing the frozen reactants (–273 °C) directly into 70 °C water, maintain the temperature for 40 minutes and simultaneously pump out the released hydrogen fluoride. Then allow the reaction vessel to cool to room temperature and wash off the remaining H$_2$SO$_4$ with a large portion of anhydrous HF. The result of powder XRD measurements of our best sample (sample **1**), performed at RT, is shown in **Figure 1**. The obtained unit cell was found to be a derivative of the CuSO$_4$-type orthorhombic structure but with the *β* deviating from 90°; thus, we based the crystal structure refinement on atomic coordinates from CuSO$_4$.[14] The β-AgSO$_4$ crystallizes in the monoclinic system ($P2_1/n$), with $a$ = 5.1189(2) Å, $b$ = 8.7489(13) Å, $c$ = 7.2023(10) Å, $β$ = 94.104(9)°, and $V$ = 298.00(7) Å$^3$ for $Z$ = 4 at room temperature. The non-standard setting $P2_1/n$ for the No. 14 space group was chosen here to keep similar bonding topology as in the parent CuSO$_4$ structure. The unit cell volume per one formula unit

[a] M. Domański, Prof. W. Grochala
Center of New Technologies
University of Warsaw
Zwirki i Wigury 93, 02-089 Warsaw Poland
E-mail: m.domanski@cent.uw.edu.pl
[b] Dr Zoran Mazej
Department of Inorganic Chemistry and Technology
Jožef Stefan Institute
Jamova cesta 39, SI-1000 Ljubljana Slovenia

Supporting information for this article is given via a link at the end of the document.

(*ca.* 74.50 Å$^3$) is about 1.8% smaller than the volume of α-AgSO$_4$. Further details of the crystal structure reported here may be obtained from CCDC/FIZ Karlsruhe on quoting the CSD deposition No. 2191789 (*cf.* also **Table S1** in **Supporting Information**).

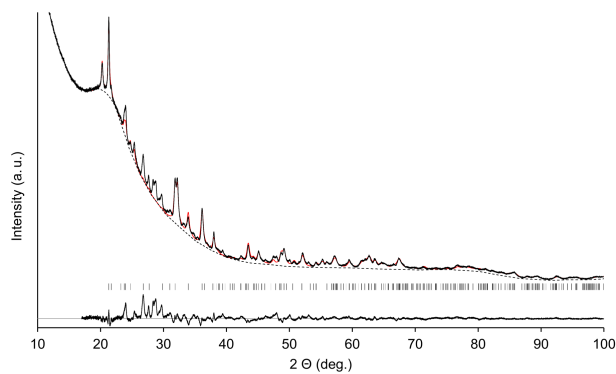

**Figure 1.** Measured PXRD pattern of our best sample of β-AgSO$_4$ (black) and the calculated pattern (red), together with the manual background (dashed line). The differential profile is at the bottom.

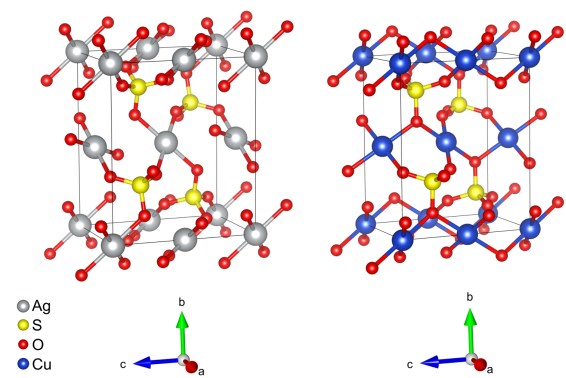

**Figure 2.** View of the crystal structures of experimentally obtained β-AgSO$_4$ in *P*2$_1$/*n* setting and CuSO$_4$. The threshold for Ag-O and Cu-O bond drawing is set to 2.3 Å. Note the absence of any direct Ag-O-Ag linker in the former and its presence for Cu analogue.

Some discrepancies between the measured pattern and that generated based on our model are present, as seen in the difference pattern especially in the range 22-30° of 2Θ. However, no crystalline impurities containing Ag, S, O, H and/or F, could be unambiguously assigned to these reflections. Possibility that the unit cell vectors are correct but symmetry is lower than proposed, should be excluded, as this is contradicted by the lack of possible reflections in that 2Θ range even for the *P*1 space group. However, it cannot be ruled out that β-AgSO$_4$ has an even more complex structure than the one proposed here, *e.g.* there is a very small symmetry-lowering distortion which leads to a supercell, or it has a modulated structure along one or more crystallographic directions. Nevertheless, as we will see, the monoclinic cell discussed here is a reasonable model, which grasps all key structural features.

The most pronounced difference between β-AgSO$_4$ and CuSO$_4$ is the loss of orthorhombic symmetry in the case of β-AgSO$_4$. Why the monoclinic distortion occurs will be further investigated in the theoretical DFT calculations section. Still, these structures show many similarities (**Figure 2**). In both we can find that the shortest metal-metal distances are along the [001] direction, and the metal cations adopt a distorted octahedral coordination with an axial bonds elongation (*i.e.* tetragonal elongation). This means that unpaired electrons sit on d($x^2$–$y^2$) type of orbitals which are placed within [MO$_4$] square planar units. The most important structural difference between β-AgSO$_4$ and CuSO$_4$ is in the interatomic connectivity inside the unit cell. In the copper sulfate, some links between Cu sites go via oxygen atoms bridges, which means that [CuO$_4$] units are connected with apices. This means that the strongest spin-spin interaction should occur along [001] chains. In the silver sulfate, it is not the case since [AgO$_4$] units are rotated in such a way that one O···Ag separation is long (2.571(9) Å) and spin-spin interactions must be weak along [001] direction.

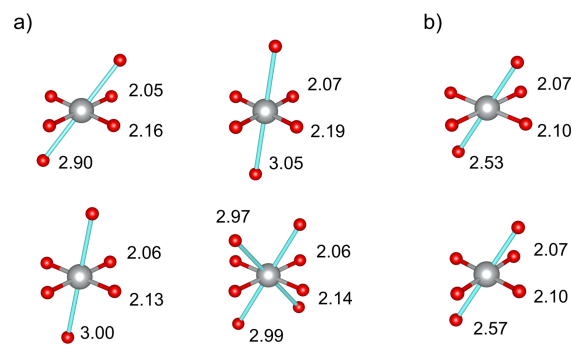

**Figure 3.** Comparison of coordination spheres for Ag sites in (a) α-AgSO$_4$ and (b) β-AgSO$_4$. Bond lengths are given in Å. The threshold for Ag-O bonds drawing is set to 3.1 Å, but longer apical bonds are marked in cyan colour.

In the terms of the first coordination sphere of silver, β-AgSO$_4$ differs also from the previously known α-AgSO$_4$. The silver site coordinations in both polymorphs are compared in **Figure 3**. In the new polymorph, we find only two crystallographically different Ag sites, while in α-AgSO$_4$ there are four. In both polytypes, the four shortest Ag-O bonds are significantly shorter than the rest, meaning that in both compounds the Ag d$^9$ hole is located on d($x^2$–$y^2$) type orbitals parallel to these square planar units. However, in β-AgSO$_4$ apical oxygen atoms are located much closer to the silver atoms, explaining a slight reduction of unit cell volume compared to that of α-AgSO$_4$ (1.8% decrease). Moreover, the shortest Ag-Ag distances also differ considerably: 4.68 Å in α and 3.60 Å in β polymorph. Last but not the least, all [AgO$_6$] octahedra in β-AgSO$_4$ exhibit strong trigonal distortion (the lowest torsion angles are 56.1° and 62.3° for Ag1 and Ag2 sites, respectively), while similar coordination is seen only for one Ag site in α-AgSO$_4$ (the respective angle is 50.6°). Despite these differences, α-AgSO$_4$ and β-AgSO$_4$ show similarity in the topology of layers perpendicular to the lattice vector *b* (**Figures S1-S2**).

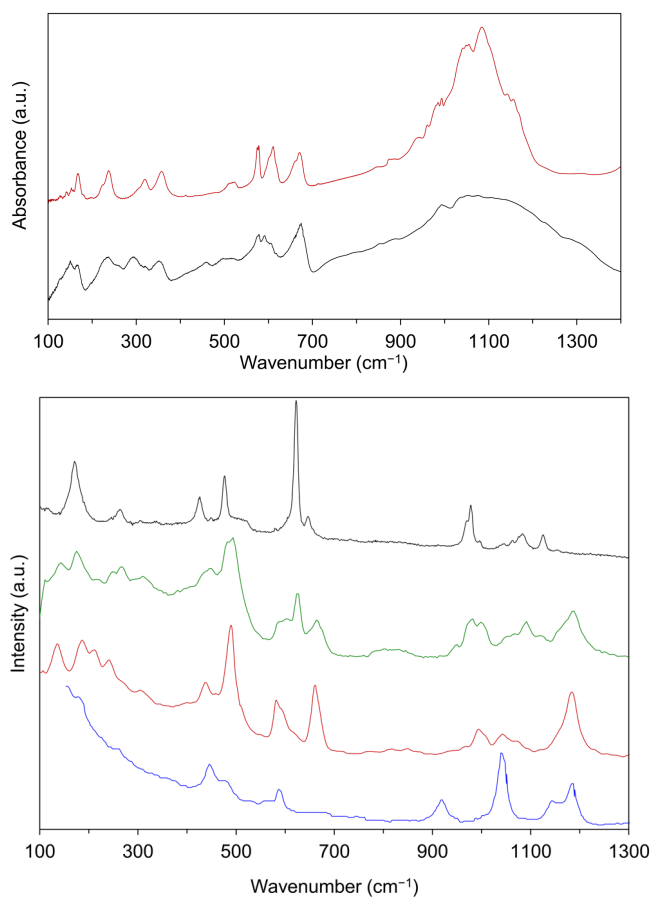

**Figure 4.** (top) The combined FIR and MIR spectra of sample **2** (black) and the previously published α-AgSO$_4$[10] (red). The absorption hump from 900 to 1300 cm$^{-1}$ in sample **2** is broader than for α-AgSO$_4$. (bottom) Raman spectra of the previously published α-AgSO$_4$ (black curve, 647 nm Ar-Kr laser[16], electro-synthesized product), the sample **2** (green curve, 532.8 nm Nd:YAG laser), the sample **3** (red curve, 632.8 nm He–Ne laser), and CuSO$_4$ (blue curve, 514.5 nm Ar laser[19]).

**Vibrational spectra**

Novel Ag(II) compounds as well as most metal sulphates have been previously characterized employing vibrational spectroscopy [10,12, 13,16-18]. However, the amount of sample **1** allowed only for performing powder XRD measurements, and we faced difficulties in repeating synthesis. For this reason, spectral analyses were performed on sample **2**, which consisted of 58.2(2)% β-AgSO$_4$, and 41.8(2)% α phase, as seen from Rietveld analysis (PXRD in **Figure S3**). The combined MIR and FIR spectra of β-AgSO$_4$ and α-AgSO$_4$ are shown in **Figure 4**. We notice key differences in the IR spectrum of sample **2** as compared to α-AgSO$_4$: (1) appearance of new bands which may be assigned to β-AgSO$_4$ (150, 259, 295, 352, 459, 498, 591, 606, 674 cm$^{-1}$), (2) the presence of a broader but relatively less intense hump from 900 to 1300 cm$^{-1}$ originating from S-O bond stretching vibrations, and (3) the fact that bands of sample **2** are wider than those for α-AgSO$_4$ which may come from the fact that **2** is a mixture of two forms.

The measured Raman spectra of sample **2**, α-AgSO$_4$, and CuSO$_4$ are shown in **Figure 4**. Raman spectrum of both anhydrous copper sulphate[19] and α-AgSO$_4$[16] show some similarities with the spectrum of sample **2**. Still, there are many unique bands in the spectrum of sample **2** which cannot be assigned to α-AgSO$_4$. This is consistent with the PXRD analysis of the sample **2** which indicates the presence of both polymorphs. We also collected the Raman spectrum of independently prepared sample **3** (**Figure 4**). In this spectrum new features are very much enhanced, and bands of α-AgSO$_4$ are absent. Thus, Raman spectrum of sample **3** may be treated as genuine fingerprint of β-AgSO$_4$. Regretfully, the amount of sample **3** was insufficient to confirm its high purity with PXRD.

**Table 1.** New bands observed in the vibrational spectra of β-AgSO$_4$ (samples **2** and **3**) together with DFT+U calculated wavenumbers, symmetry and their assignment.

| $\tilde{\nu}$ (IR) [cm$^{-1}$] | $\tilde{\nu}$ (R) [cm$^{-1}$] | DFT [cm$^{-1}$] | Symmetry | Assignment |
|---|---|---|---|---|
| | 136 s | 142 | Ag | r [SO$_4$]$^{2-}$ |
| 150 m | | 148 | Au | r [SO$_4$]$^{2-}$ |
| | 187 m | 192 | Bg | $\nu_1$ [Ag-O$_{apical}$] + r [SO$_4$]$^{2-}$ |
| | 212 m | 212 | Ag | $\nu_1$ [Ag-O] + r [SO$_4$]$^{2-}$ |
| | 242 m | 231 | Bg | $\nu_1$ [Ag-O] + r [SO$_4$]$^{2-}$ |
| 259 w | | 278 | Au | $\nu_2$ [Ag-O] + t [SO$_4$]$^{2-}$ |
| 295 m | | 291 | Bu | $\nu_2$ [Ag-O] + t [SO$_4$]$^{2-}$ |
| | 304 m | 303 | Bg | $\nu_1$ [Ag-O] + r [SO$_4$]$^{2-}$ |
| 321 w | | - | - | - |
| 352 m | | 353 | Au | $\nu_2$ [Ag-O] + r [SO$_4$]$^{2-}$ |
| | 438 m | 445 | Bg | b$_1$ [SO$_4$]$^{2-}$ + b$_4$ [SO$_4$]$^{2-}$ |
| 459 m | | 457 | Au | b$_1$ [SO$_4$]$^{2-}$ + b$_4$ [SO$_4$]$^{2-}$ |
| | 490 vs | 484 | Ag | b$_1$ [SO$_4$]$^{2-}$ |
| 498 w | | 503 | Au | b$_1$ [SO$_4$]$^{2-}$ |
| | 582 s | 582 | Bg | b$_3$ [SO$_4$]$^{2-}$ |
| 591 m | | 581 | Bu | b$_2$ [SO$_4$]$^{2-}$ |
| | 596 sh | 598 | Bg | b$_2$ [SO$_4$]$^{2-}$ |
| 606 sh | | 606 | Bu | b$_2$ [SO$_4$]$^{2-}$ |
| | 618 sh | 599 | Ag | b$_2$ [SO$_4$]$^{2-}$ |
| | 661 vs | 666 | Ag | b$_3$ [SO$_4$]$^{2-}$ |
| 674 vs | | 675 | Bu | b$_3$ [SO$_4$]$^{2-}$ |
| | 849 w | - | - | - |
| 994 vs | | 1009 | Au | $\nu_3$ (S-O) |
| | 995 s | 1012 | Bg | $\nu_1$ (S-O) |
| | 1043 m | 1053 | Ag | $\nu_3$ (S-O) |
| | 1072 sh | 1072 | Bg | $\nu_3$ (S-O) |
| | 1158 sh | 1158 | Bg | $\nu_2$ (S-O) |
| | 1184 vs | 1180 | Ag | $\nu_2$ (S-O) |
| 900 – 1300 vs, br | | 1038, 1097 | Au | $\nu_3$ (S-O), $\nu_1$ (S-O) |
| | | 1049, 1088, 1133 | Bu | $\nu_3$ (S-O), $\nu_1$ (S-O), $\nu_2$ (S-O) |

Intensity code: vs very, s strong, m medium, w weak, sh shoulder, br broad. Assignment code: $\nu_1$ – symmetric stretching, $\nu_2$ – asymmetric stretching, $\nu_3$ – umbrella stretching, b$_1$ – bending (in-phase), b$_2$ – bending (in counter-phase), b$_3$ – wagging, b$_4$ – twisting, r – hindered rotation, t – hindered translation. Vibrational and bending modes are schematically shown in **Figure S7**.

We notice that Raman spectroscopy is much better tool than FTIR to distinguish between the two polymorphs of AgSO$_4$. It is also worth emphasizing that, just like α-AgSO$_4$, the new polymorph is photosensitive, and it decomposes in a strong laser beam (**Figure S4**). However, neither cooling down to 77 K nor warming up to 343 K changed significantly the Raman spectra of sample **3** (**Figure S5**). This suggests that β-AgSO$_4$ is stable or at least metastable at this temperature range.

Having detected bands which are unique to β-AgSO$_4$, we have also computed vibrational fundamentals (phonons) using the DFT+U approach. The obtained wavenumbers and symmetries for respective modes are presented in **Table S2**. Summary of bands positions, the DFT-calculated values of phonon wavenumbers at Γ-point and assignment of bands to the modes are presented in **Table 1**. The correlation between experimental and theoretical values is excellent ($R^2$=0.9998) as shown in **Figure S6**.

### Stability, magnetic superexchange and electronic band gap as viewed from theoretical DFT calculations

Magnetic structure of CuSO$_4$ was investigated intensely already in the 1960s, both theoretically[20] and experimentally[14]. It was noted that there are 3 main spin-spin interactions (called therein $J_0$, $J$ and $J'$), but involving 5 different superexchange paths. This is the case for orthorhombic crystal structure, but in the monoclinic system of related β-AgSO$_4$ these superexchange constants may begin to differentiate for each superexchange path.

**Table 2.** Unit cell volumes (in Å$^3$), band gap energy ($E_{bg}$, in eV), relative electronic energies (in kJ/mol) of groundstate ($\Delta E_{GS}$) and metallic state ($\Delta E_{met}$) and ZPE-corrected energies of groundstates ($\Delta E^{ZPE}$) calculated for α-AgSO$_4$ and β-AgSO$_4$ structures of different space groups (SGs) with two different DFT approaches.

| Phase | SG | Method | V | ΔV | $E_{bg}$ | $\Delta E_{GS}$ | $\Delta E_{met}$[a] | $\Delta E^{ZPE}$ |
|---|---|---|---|---|---|---|---|---|
| α | C2/c | DFT+U | 78.29 | 0 | 1.10 | 0 | 0 | 0 |
| | C2/c | HSE06 | 77.72 | 0 | 2.02 | 0 | | |
| β | Pnma | DFT+U | 78.21 | -0.08 | 0.90 | -0.28 | -2.60 | -1.01 |
| | P2$_1$/n | DFT+U | 79.36 | 1.06 | 0.81 | -0.33 | 1.50 | -1.06 |
| | P2$_1$/n [b] | DFT+U | 79.21 | 0.92 | 0.85 | -0.65 | 1.46 | -1.36 |
| | Pnma | HSE06 | 77.79 | 0.06 | 1.93 | 1.78 | | |
| | P2$_1$/n | HSE06 | 79.05 | 1.33 | 1.74 | 0.61 | | |
| | P2$_1$/n [b] | HSE06 | 79.32 | 1.60 | 1.68 | 0.24 | | |

[a] $\Delta E_{met}$ are relative energies for non-spin-polarized single-point calculations.
[b] The 2x1x1 monoclinic supercell allows for reproduction of the lowest magnetic ground state, after the geometry optimization.

Theoretical methods were used here to investigate (1) the relative energy of β-AgSO$_4$ with respect to α-AgSO$_4$ and (2) why the monoclinic distortion occurs in the structure of β-AgSO$_4$. For the latter purpose, both the hypothetical undistorted orthorhombic (Pnma) and monoclinic (P2$_1$/n) versions of β polymorph were computed with the DFT methods. The key geometric features of β-AgSO$_4$ crystal structure where confirmed by theory; e.g., the β angle remains >90° (exp. 94.1°, theor. 95.2°). Also, there are two independent Ag sites, differing in coordination environment and magnetization value (±0.494 μ$_B$ for Ag1 and ±0.534 μ$_B$ for Ag2). Some discrepancies between experimental and DFT optimized structures exist (**Figure S8**), e.g. the local trigonal distortions at the Ag(II) sites are noticeably weaker for the latter.

A summary of DFT results is presented in **Table 2**. Experimentally, β polymorph has a slightly smaller unit cell volume than α-AgSO$_4$ (–1.38 Å$^3$ per formula unit), but this subtle difference is not reproduced by DFT calculations. The electronic energy relation between the two polymorphs is also unclear – on the DFT+U level the β polymorph has a slightly lower electronic energy, but with hybrid functional HSE06 it is the opposite. Thus, relative stability of both polytypes is unclear at present. It is important to remark that we tend to call β-AgSO$_4$ as the high-temperature form, but this is solely due to the fact that it has been obtained via a transition from α-AgSO$_4$ at elevated temperature.

The vibrational zero-point energy (ZPE) was calculated at relatively inexpensive DFT+U level for both forms. Calculations indicate that β-AgSO$_4$ has significantly lower ZPE than α-AgSO$_4$. This should give β-AgSO$_4$ some preference over α-form at elevated temperature (**Figure S9**) due to a larger vibrational entropy term, as seen also in experiment.

**Table 3.** Summary of superexchange-mediated spin-spin interactions in the 2x1x1 supercell of β-AgSO$_4$ obtained with HSE06. Type *o* and *m* mean "one-site" (i.e. Ag1-Ag1 and Ag2-Ag2) and "mixed" (i.e. Ag1-Ag2) type of spin interactions.

| | Type | Direction | Ag'-O (Å) | O-Ag" (Å) | Ag'-Ag" (Å) | ∠Ag'-S-Ag" (°) | J (meV) |
|---|---|---|---|---|---|---|---|
| J$_1$ | m | [001] | 2.102 / 2.127 | 2.876 / 2.527 | 3.673 | 94.6 /106.2 | 5.9 |
| J$_{2\_1}$ | o | [100] | 2.111 | 2.878 | 4.841 | 83.8 | 1.4 |
| J$_{2\_2}$ | o | [100] | 2.109 | 2.123 | 4.841 | 95.0 | -20.4 |
| J$_3$ | m | (001) | 2.102 | 2.108 | 5.083 | 102.3 | -4.6 |
| J$_4$ | m | [101] | 2.111 | 2.123 | 5.809 | 120.9 | -38.9 |
| J$_{5\_1}$ | o | (201) | 2.102 | 2.110 | 6.143 | 135.2 | -38.1 |
| J$_{5\_2}$ | o | (201) | 2.123 | 2.526 | 6.143 | 127.3 | 2.1 |

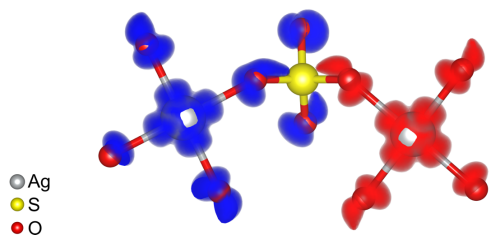

○ Ag
○ S
● O

**Figure 5.** The spin-density along J$_4$ interaction in HSE06 structure of β-AgSO$_4$. Notice a significant amount of spin-density of opposite signs (red and blue) on sulfate's oxygen atoms facilitating the superexchange, and lack of spin on sulphur atom. The isosurface value is 0.005 e/Å$^3$.

One important result is related to a hypothetical undistorted orthorhombic polymorph, Pnma, having direct Ag-O-Ag linkers. Using the spin non-polarized method (that corresponds to a metallic and non-magnetic scenario), the orthorhombic β-AgSO$_4$ is preferred over the monoclinic β-AgSO$_4$ form by as much as 4.1 kJ/mol. However, when spin-polarization is included, with both DFT functionals the monoclinic β-AgSO$_4$ has lower energy than the orthorhombic one. This suggests that monoclinic distortion is largely due to magnetic interactions between Ag(II) sites. This result implies large magnetostrictive coupling and it emphasizes the importance of spin degree of freedom for correct description of crystal structure of Ag(II) compounds.

But just how strong are these magnetic interactions? Here, diverse pathways of magnetic interactions in a monoclinic β-AgSO$_4$ were considered and superexchange coupling constants were calculated using the Heisenberg Hamiltonian of the type $H = -\sum_{ij} J_{ij} s_i s_j$ with ($J_{ij} \neq J_{ji}$). The resulting coupling constants are presented in **Table 3**, while topological and computational details are presented in **Figures S10-S11** and **Tables S3-S5**. Our calculations predict a very strong antiferromagnetic interaction $J_4$ – along a superexchange pathway along *ac* plane diagonal (pictured in **Figure 5**), which is allowed only by the monoclinic distortion. In fact, the corresponding magnetic superexchange constant of *ca.* −39 meV is probably record strong among all known chemical compounds which host superechage pathway encompassing as many as four bonds (here, Ag–O–S–O–Ag). Strong AFM interactions were initially computed also along *a* lattice vector, which indicated that the construction of the 2x1x1 supercell might lead to an even lower electronic energy. That was confirmed via optimization of the 2x1x1 supercell (see **Table 2**), however no symmetry reduction was observed. Regretfully, we have not been able to experimentally determine magnetic properties of β-AgSO$_4$ due to the lack of a specimen which would be sufficiently free from the α-form. However, theory clearly shows that spin polarization omits S atom (**Figure 5**), just like in the α-form.[10]

The last important result stemming from theoretical calculations is the comparison of the fundamental electronic bandgap values. Both functionals consistently predict smaller gap for β-AgSO$_4$ than α-AgSO$_4$ (DFT+U: 0.85 eV *vs.* 1.10 eV; HSE06: 1.67 eV *vs.* 2.02 eV). The experimental values await to be determined.

### Thermal stability

The thermogravimetric and differential scanning calorimetry profiles, and ionic current for O$_2$ ions are presented in **Figure 6**. During the heating of sample **2**, two oxygen waves were observed. The first wave started at about 120 °C and the second at about 220 °C, indicating that sample **2** indeed contains both low-temperature and high-temperature phases of AgSO$_4$ in agreement with the PXRD analysis. Note, α-AgSO$_4$ decomposes above 120 °C according to the reaction equation[10]:

$$2\ AgSO_4 \rightarrow Ag_2S_2O_7 + \tfrac{1}{2} O_2$$

Thus, the exothermic event at 220 °C may be attributed to the thermal decomposition of β-AgSO$_4$. The observed mass loss up to 260 °C is −4.22% (−1.92% for the first step of decomposition and −2.30% for the second step), not far from the theoretical values (−3.92% total, −1.64% first step, −2.28% second step for the phase ratio of sample **2** as results from PXRD).

It is also worth noting that thermal decomposition profile for sample **2** which contains both polymorphs is not a simple sum of the two decomposition profiles. Notably, (1) heating of sample **2** above 250 °C results in only a 10.1% mass decrease until reaching 500 °C, compared to 22.1% for the pure α-AgSO$_4$ (theoretical value 23.5%). This suggests that much more SO$_3$ is now kept within solid residue, possibly in the form of a polysulphate. (2) Another difference is apparent in DSC profiles, as Ag$_2$S$_2$O$_7$ intermediate decomposes endothermally with a peak around 350 °C while here we see a endothermal peak at temperature above 400 °C. (3) Also, PXRD pattern analysis of the solid residue suggests that as-yet unknown product of thermal decomposition has formed (possible a said polysulphate), suggesting a complex mechanism of decomposition for sample **2**.

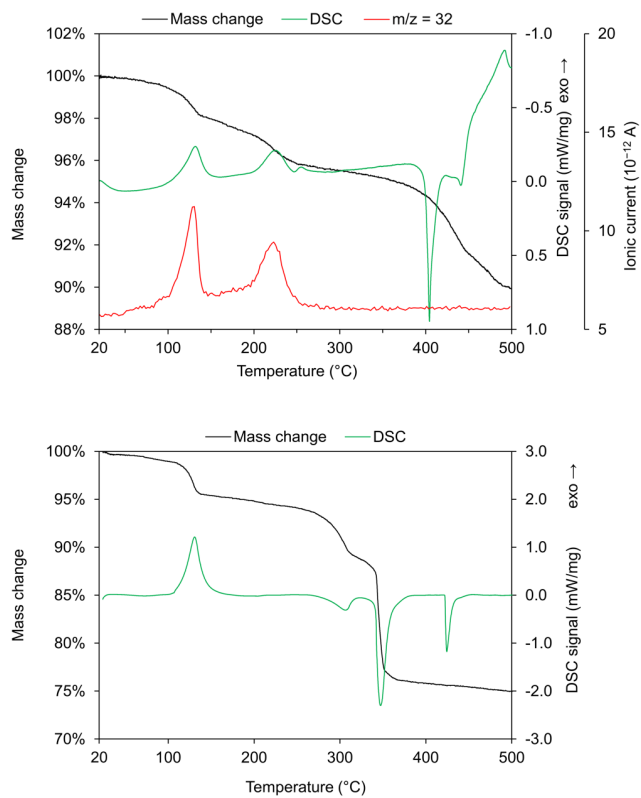

**Figure 6. Top:** Thermal gravimetry, differential scanning calorimetry, and ionic current for oxygen ions (m/z=32) signals collected for sample **2** of AgSO$_4$. Heating rate 5 K/min. **Bottom:** Thermal gravimetry curve and differential scanning calorimetry signal collected for the α-AgSO$_4$ (reprint from Ref.[4]).

## Conclusions

We have described the new crystalline form of AgSO$_4$ whose structure is related to that of CuSO$_4$ but it lacks direct Ag-O-Ag links. The β-AgSO$_4$ unit cell shows monoclinic distortion from the parent orthorhombic type, which was found to appear due to strong antiferromagnetic interactions along the *ac* diagonal. The strength of this interaction is close to −39 meV, which is record large for any superexchange pathway involving three-atomic bridge, such as O-S-O linker here. Moreover, spin polarization omits the Sulphur atom and is realized by unusual O···O interaction. The IR spectra of both AgSO$_4$ polymorphs are similar, but large differences in their Raman spectra may be used to handily distinguish between the two forms. β-AgSO$_4$ is more thermally stable from its α-predecessor, with the onset for oxygen evolution higher by 100 °C than that for the α-form. The potential application of the new form as oxidizer in organic chemistry will soon be explored.

## Experimental Section

Synthesis: $AgF_2$ (0.4 g) was loaded into a reaction vessel in a dry-box. On a vacuum system, $AgF_2$ was cooled to 77 K and liquid anhydrous $H_2SO_4$ (2 ml) was poured onto it (**Figure S12**). The frozen reaction vessel was then placed in hot water (70 °C) for 40 min and HF was pumped off simultaneously. The reaction mixture was then cooled to RT and a large amount of HF was added to wash away all soluble phases.

The powder XRD measurements were conducted at RT using a Malvern Panalytical Empyrean diffractometer with a Cu Kα wavelength, no monochromator and an RTMS detector. After indexation of the measured XRD reflections, the unit cell was found using X-Cell[21] (available in the Materials Studio suite). The Rietveld method was applied to the measured pattern to determine the consistency of the crystal structure model sample **1** and verify the crystal phases ratio for sample **2**. The crystal structure of β-$AgSO_4$ was solved in Jana2006[22]. Due to the complex shape of the amorphous background, manual background correction was used. Constraints on the structure parameters' were applied, details are listed in the provided CIF file. VESTA[23] was used for structure visualization.

Thermal analysis of sample **2** was performed using a thermogravimeter combined with a differential scanning calorimeter from Netzsch STA 409 PG (NETZSCH-Gerätebau GmbH, Selb, Germany) using $Al_2O_3$ crucibles. The oven was constantly purged at a constant Ar (6N) flow rate of 80 mL/min$^{-1}$. The evolved gases were analysed with a quadrupole mass spectrometer (MS) QMS 403 C (Pfeiffer Vacuum, Aßlar, Germany), connected to the TGA/DSC device by a quartz capillary preheated to 200 °C to avoid condensation of low-boiling volatiles. The heating rate was set to 5 K/min.

Fourier-transform infrared spectroscopy (FTIR) measurements of sample **2** were carried out using a Bruker Vertex 80V spectrometer (Bruker, Billerica, MA, USA). For FIR measurements, pure fine powdered sample **2** was placed on HDPE windows; in the case of the MIR measurement anhydrous KBr stored in a dry-box was used as a pellet material dispersing sample **2**.

Raman spectra were collected with Horiba Jobin Yvon Raman spectrometer with 532 nm Nd:YAG laser exciting beam for sample **2** (**Figure 4**) and with Horiba Jobin Yvon LabRam-HR Raman microspectrometer with 632.8 nm He–Ne laser exciting beam for sample **3** (presented in **Figure 4** and **Figures S3-4**).

Spin-polarized DFT calculations were performed using generalized gradient approximation exchange-correlation functional PBEsol[24] with projected augmented wave method[25] as implemented in Vienna Ab initio Simulation Package 5.4.4 (VASP)[26]. The plane-wave energy cut-off was set for 600 eV, and k-spacing to 0.25 Å$^{-1}$ for ionic optimization and 0.30 Å$^{-1}$ for phonon calculations, while the threshold of $10^{-7}$ eV for electronic self-consistent-field convergence. We used DFT+U formalism to incorporate the strong on-site interactions characteristic for open-shell transition metals I. The rotationally invariant approach as proposed by Liechtenstein[27] was used. The on-site interactions were included for each of the elements with the Hubbard term $U$ set 5 eV for Ag ($l = 2$), 2 eV for S ($l = 1$) and 4 eV for O ($l = 1$), and exchange term $J$ set equally for all the atoms (1 eV). A similar computational setup can be found in other studies of silver(II) sulphate and sulphate hydrate[11,13,16]. For the spin non-polarized DFT single-point calculations the k-spacing was set to 0.15 Å$^{-1}$. For the more resource-demanding spin-polarized calculations with HSE06 hybrid functional, the k-spacing was set to 0.30 Å$^{-1}$. The electronic band gaps were calculated using tetrahedron method with Blöchl corrections.

The vibrational zero-point energy (ZPE) corrections were obtained using the finite displacement method as implemented in the Phonopy software[28] using VASP as the force calculator. Before the Γ-frequencies calculations, the forces per atom were converged below the value of $10^{-4}$ eV Å$^{-1}$.

## Acknowledgements

This research was supported by the Polish National Science center (NCN) within WG's Maestro project (2017/26/A/ST5/00570). Research was carried out with the use of CePT infrastructure financed by the European Union – the European Regional Development Fund within the Operational Programme "Innovative economy" for 2007–2013 (POIG.02.02.00-14-024/08-00). ZM acknowledges the financial support of the Slovenian Research Agency (research core funding No. P1-0045; Inorganic Chemistry and Technology).

**Keywords:** silver • sulfate • oxidizers • polymorphism • magnetism

# New CuSO$_4$-related high-temperature polymorph of Ag$^{II}$SO$_4$

Mateusz Domański,*[a] Zoran Mazej,[b] and Wojciech Grochala[a]

[a]   M. Domański, Prof. W. Grochala
     Center of New Technologies
     University of Warsaw
     Żwirki i Wigury 93, 02-089 Warsaw Poland
     E-mail: m.domanski@cent.uw.edu.pl
[b]   Dr Zoran Mazej
     Department of Inorganic Chemistry and Technology
     Jožef Stefan Institute
     Jamova cesta 39, SI-1000 Ljubljana Slovenia

**Table S1.** Crystallographic data for β-AgSO$_4$ obtained from powder XRD analysis. Structure refinement details include (1) manual background; (2) absorption correction of Debye-Scherer (cylinder) type, μ$_r$ : 5.1 mm$^{−1}$; (3) U(iso) refined but equal for each atom and; (4) constrained distances at 1.485(50) Å for: S1-O1; S1-O2; S1-O3; S1-O4; (5) constrained distances at 2.10(10) Å for: Ag2-O4; Ag1-O3; Ag1-O2; Ag2-O1; (6) constrained angles at 109(10)° O1-S1-O2; O1-S1 O3; O1-S1-O4; O2-S1-O3; O2-S1-O4; O3-S1-O4.

| SG, formula units, volume | $P2_1/c$ (No. 14) | Z = 4 | 298.00(7) | |
|---|---|---|---|---|
| a, b, c [Å]: | 4.7414(7) | 8.7489(13) | 7.2023(10) | |
| α, β, γ [°] | 90.0 | 94.104(9) | 90.0 | |
| Fit parameters, radiation, 2Θ range | Rp = 5.61% | Rwp = 6.05% | Cu Kα | 10 < 2Θ < 100° |

Positions (label, x, y, z, multiplicity and Wyckoff symbol)

| | | | | |
|---|---|---|---|---|
| Ag1 | 0.000000 | 0.000000 | 0.000000 | 2a |
| Ag2 | 0.000000 | 0.000000 | 0.500000 | 2c |
| S1 | 0.4887(12) | 0.1761(7) | 0.3041(9) | 4e |
| O1 | 0.737(2) | 0.0780(14) | 0.2778(10) | 4e |
| O2 | 0.571(2) | 0.3387(8) | 0.2980(12) | 4e |
| O3 | 0.282(2) | 0.1467(11) | 0.1446(16) | 4e |
| O4 | 0.365(2) | 0.1328(15) | 0.4806(13) | 4e |

Molecular geometry

| Bond lengths [Å] | | Angles [°] | |
|---|---|---|---|
| Ag1-O3 | 2.526(9) | Ag1-O1-Ag2 | 102.6(5) |
| Ag1-O3 | 2.072(9) | Ag1-O2-Ag2 | 101.2(3) |
| Ag2-O1 | 2.079(10) | Ag1-S1-Ag2 | 65.67(11) |
| Ag2-O1 | 2.073(9) | Ag2-S1-Ag2 | 97.28(18) |
| Ag2-O4 | 2.571(9) | O2-S1-O3 | 107.8(6) |
| Ag2-O4 | 2.093(9) | O3-S1-O1 | 106.8(6) |
| S1-O1 | 1.484(9) | O4-S1-O3 | 109.9(8) |
| S1-O2 | 1.477(9) | O4-S1-O2 | 113.1(8) |
| S1-O3 | 1.471(12) | O1-S1-O2 | 109.9(9) |
| S1-O4 | 1.487(12) | O2-S1-O3 | 107.8(6) |

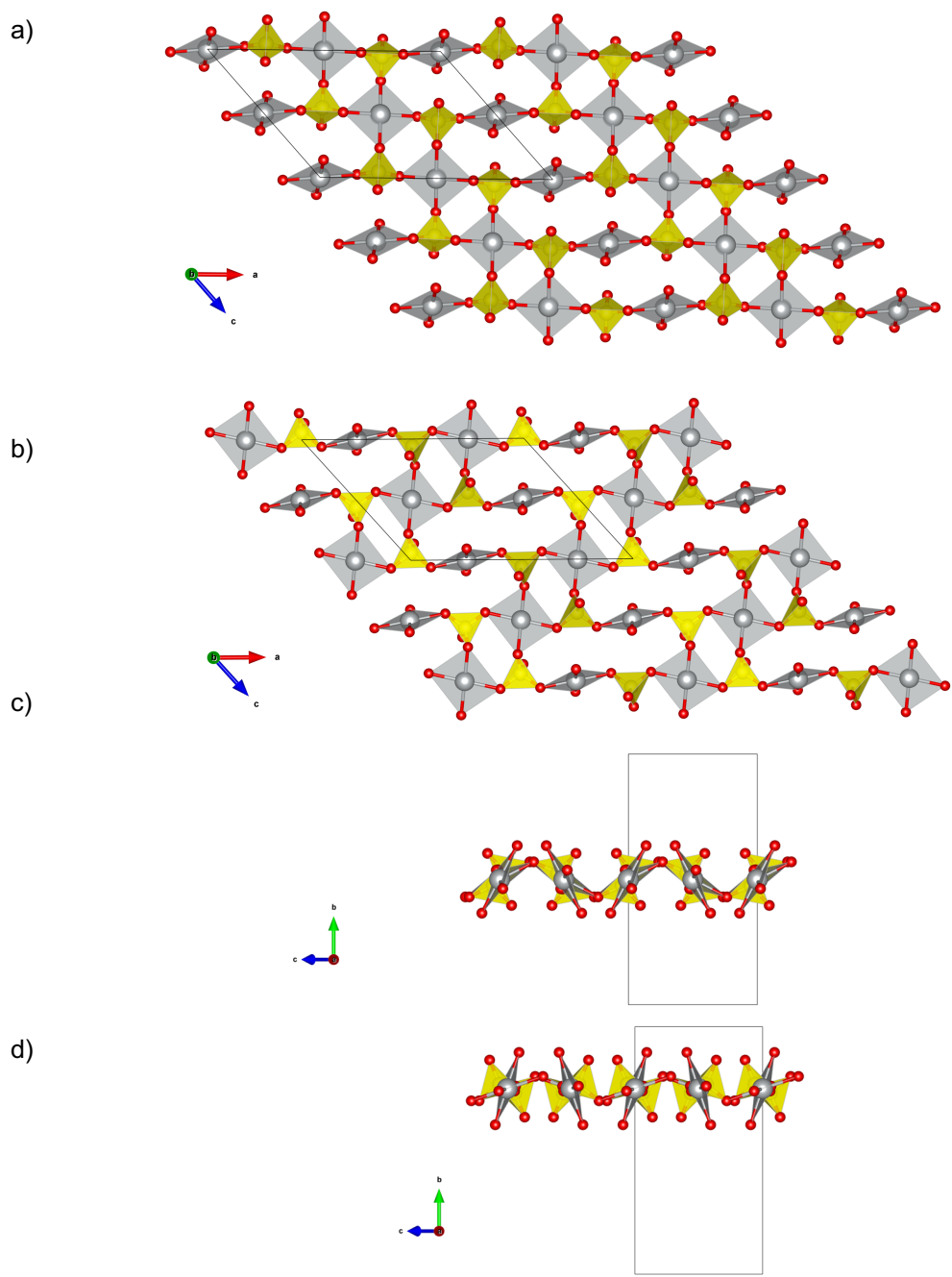

**Figure S1.** Layers perpendicular to lattice vector *b* found in the α-AgSO$_4$ crystal structure. Projections on the *ac* plane are at y = 0.5 b (a), at y = 0.75 b (b), and corresponding projections along the *a* vector (c, d).

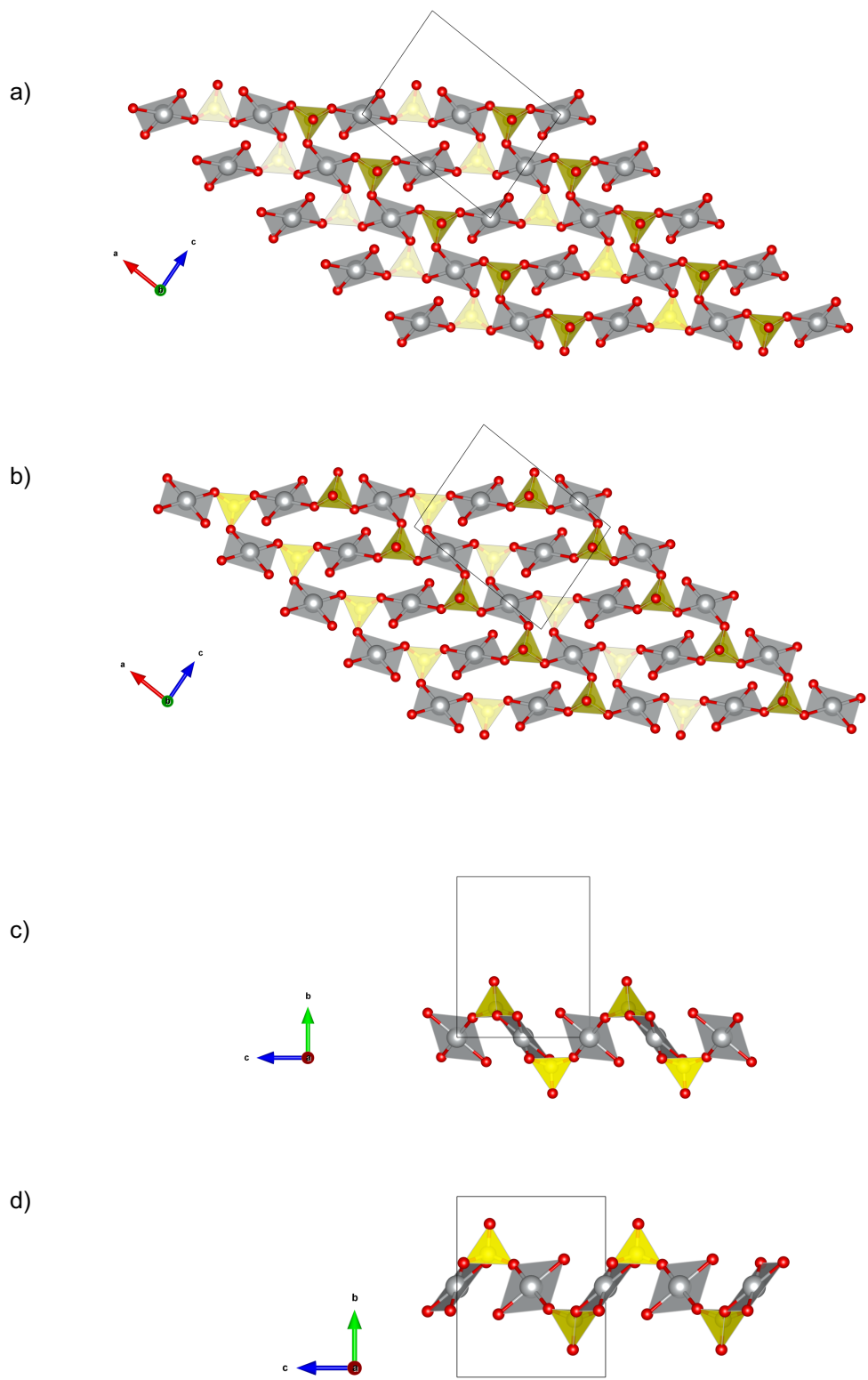

**Figure S2.** Layers perpendicular to lattice vector *b* found in the β-AgSO₄ crystal structure. Projections on the *ac* plane are at y = 0.0 b (a), at y = 0.5 b (b), and corresponding projections along the *a* vector (c, d).

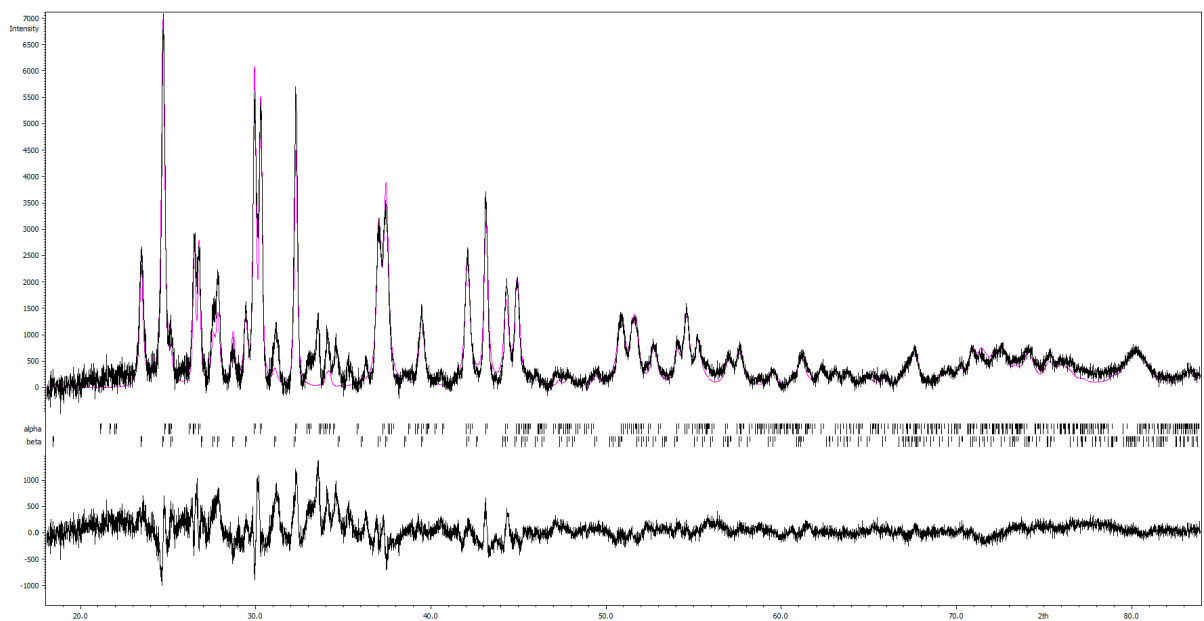

**Figure S3.** PXRD pattern of sample 2 with background subtracted and the model resulting from the Rietveld analysis (magenta) considering both polymorphs of AgSO$_4$. The Rietveld analysis shows that the content of β phase in sample 2 is about 58.2(2)%, and α phase 41.8(2)% in the mixed sample. The model does not include some peaks, what may be caused by impurities subtleties in the crystal structure of β-AgSO$_4$, similarly as in the pattern of sample 1 shown in Figure 1 in the main paper. This particular sample was later utilized to collect the Raman spectra and the thermal analyses (DSC/TGA/MS). Diffraction pattern collected radiated with a Co anode lamp, hence a relatively low signal/noise ratio.

**Table S2.** Wavenumbers (in cm$^{-1}$) and symmetry assignment of vibrational modes obtained with DFT+U computations for β-AgSO$_4$ monoclinic crystal structure.

| Raman | IR | Symmetry |
|---|---|---|
| 1195 | | Bg |
| 1180 | | Ag |
| 1158 | | Bg |
| | 1146 | Au |
| | 1134 | Bu |
| | 1098 | Au |
| | 1087 | Bu |
| 1073 | | Ag |
| 1072 | | Bg |
| 1053 | | Ag |
| | 1049 | Bu |
| | 1039 | Au |
| 1014 | | Ag |
| 1012 | | Bg |
| | 1010 | Bu |
| | 1009 | Au |
| | 688 | Au |
| | 675 | Bu |
| 675 | | Bg |
| 666 | | Ag |
| | 608 | Au |
| | 606 | Bu |
| 599 | | Ag |
| 598 | | Bg |
| 582 | | Bg |
| | 581 | Bu |
| | 580 | Au |
| 571 | | Ag |
| | 504 | Bu |
| | 503 | Au |
| 484 | | Ag |
| 478 | | Bg |
| | 458 | Bu |
| | 458 | Au |
| 445 | | Bg |
| 432 | | Ag |
| | 356 | Bu |
| | 353 | Au |
| 303 | | Bg |
| | 291 | Bu |
| 278 | | Ag |
| | 278 | Au |
| | 241 | Bu |
| | 240 | Au |
| 231 | | Bg |
| | 222 | Bu |
| | 216 | Au |
| 212 | | Ag |
| 192 | | Bg |
| 178 | | Ag |
| 162 | | Bg |
| | 161 | Bu |
| | 154 | Au |
| | 148 | Au |
| 142 | | Ag |
| | 132 | Bu |
| | 123 | Bu |
| 122 | | Bg |
| 120 | | Ag |
| | 120 | Au |
| 97 | | Ag |
| | 96 | Au |
| | 93 | Bu |
| | 88 | Au |
| | 81 | Bu |
| | 68 | Bu |
| 66 | | Bg |
| | 57 | Au |
| | 43 | Au |
| | 0 | Bu |
| | 0 | Au |
| | 0 | Bu |

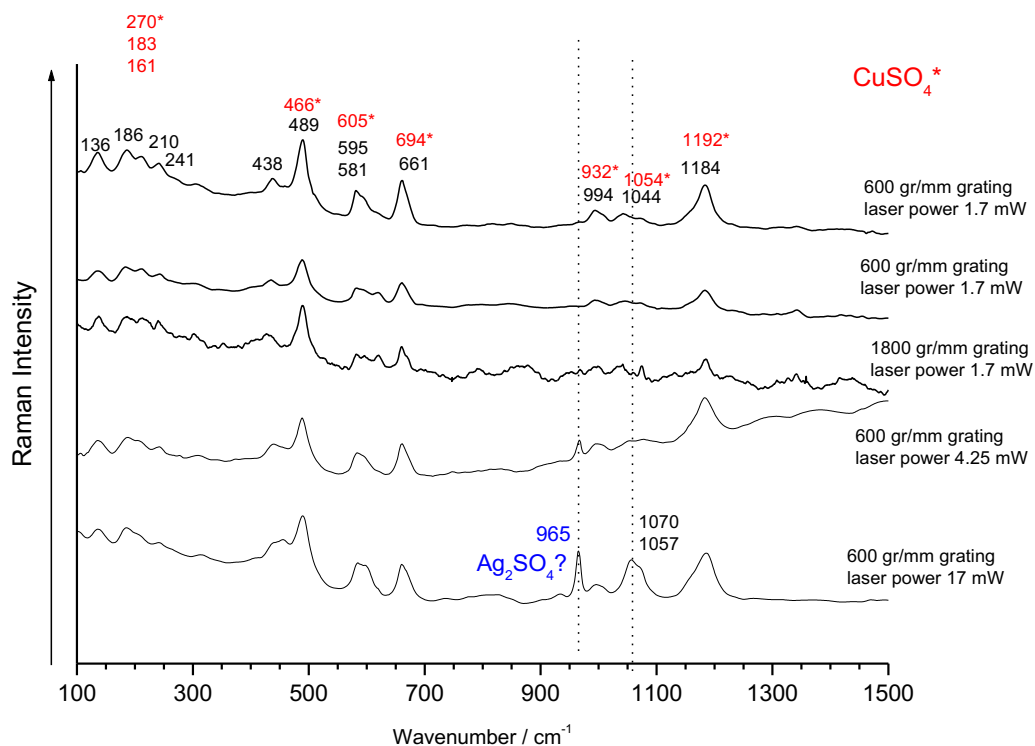

**Figure S4.** Raman spectra collected with the He-Ne red line 632.8 nm laser. Together with laser power increase the sample starts to decompose to Ag$_2$SO$_4$ and probably also Ag$_2$S$_2$O$_7$.

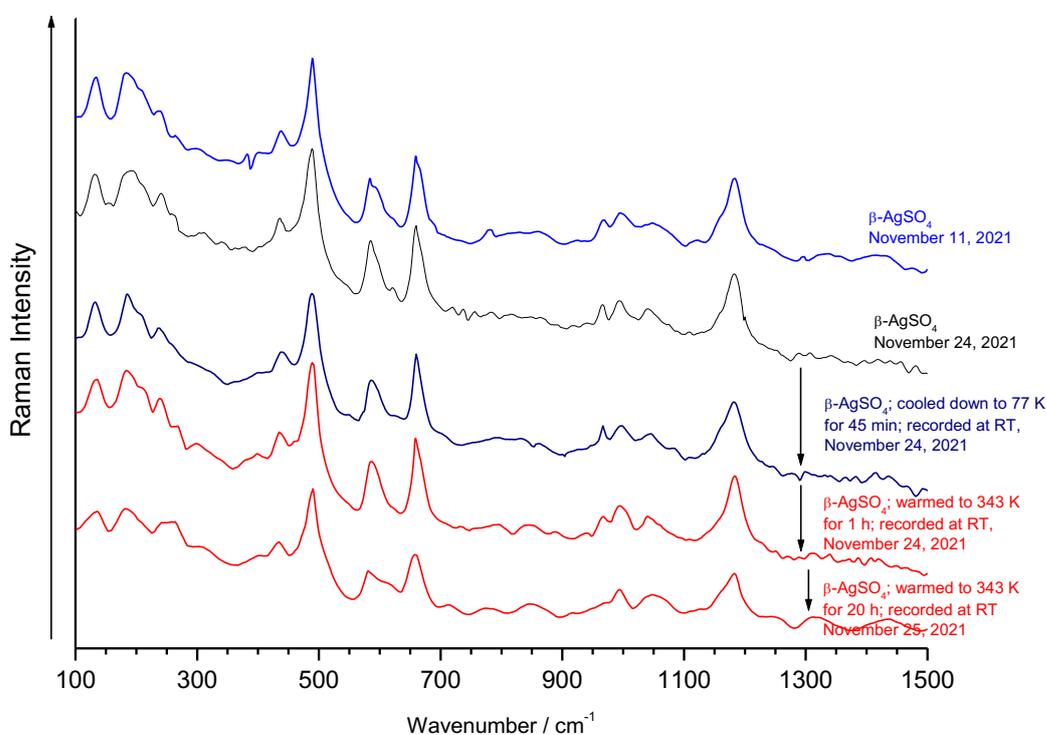

**Figure S5.** Raman spectra of the same sample (filled in a quartz capillary) collected with the He-Ne red line 632.8 nm laser. Lowering of the sample temperature does not seem to affect the collected Raman spectra, thus no temperature-driven phase transition is expected.

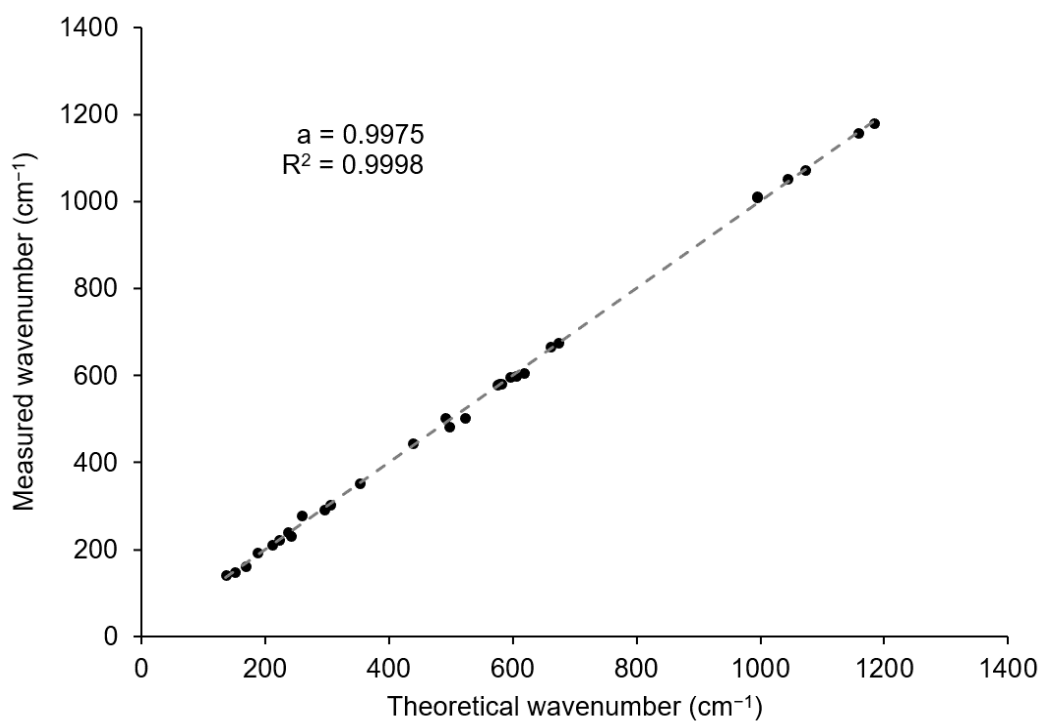

**Figure S6.** The correlation between experimental and theoretical values obtained with DFT+U approach for the modes presented in the Table 1 in the main text.

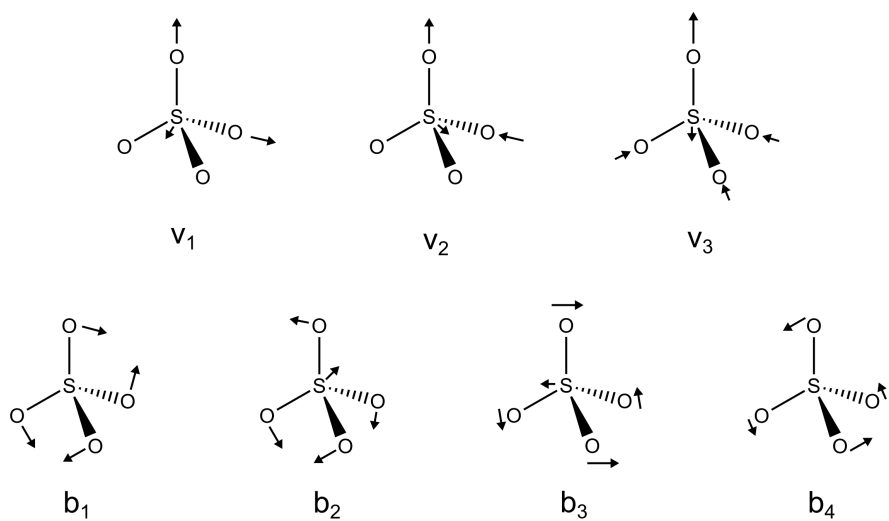

**Figure S7.** Scheme of basic stretching and bending modes of $SO_4^{2-}$ groups in β-AgSO$_4$ as resulted from DFT computations. Assignment code: $v_1$ – symmetric stretching, $v_2$ – asymmetric stretching, $v_3$ – umbrella stretching, $b_1$ – bending (in phase), $b_2$ – bending (in counter-phase), $b_3$ – wagging, $b_4$ – twisting.

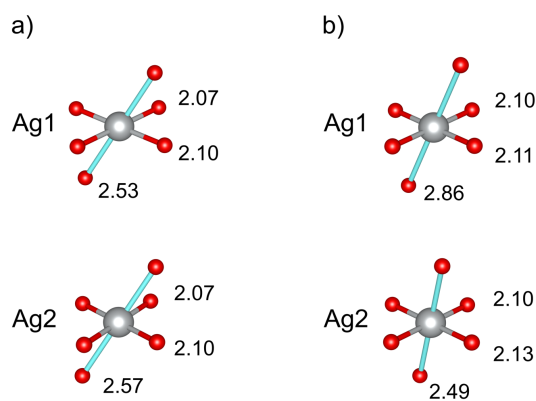

**Figure S8.** Comparison of coordination spheres for (a) Ag sites from β-AgSO$_4$ experimental crystal structure and (b) Ag sites from DFT+U optimized β-AgSO$_4$. Note that trigonal distortion in DFT+U optimized structure is much weaker – the minimal torsion angles are 71.9° for Ag1 and 80.1° for Ag2 coordination spheres, comparing to 56.1° for Ag1 and 62.3° for Ag2 sites in the experimental structure of β-AgSO$_4$. Bond lengths are provided in Å. The threshold for Ag-O bonds drawing is set to 3.1 Å, but apical bonds are marked with cyan.

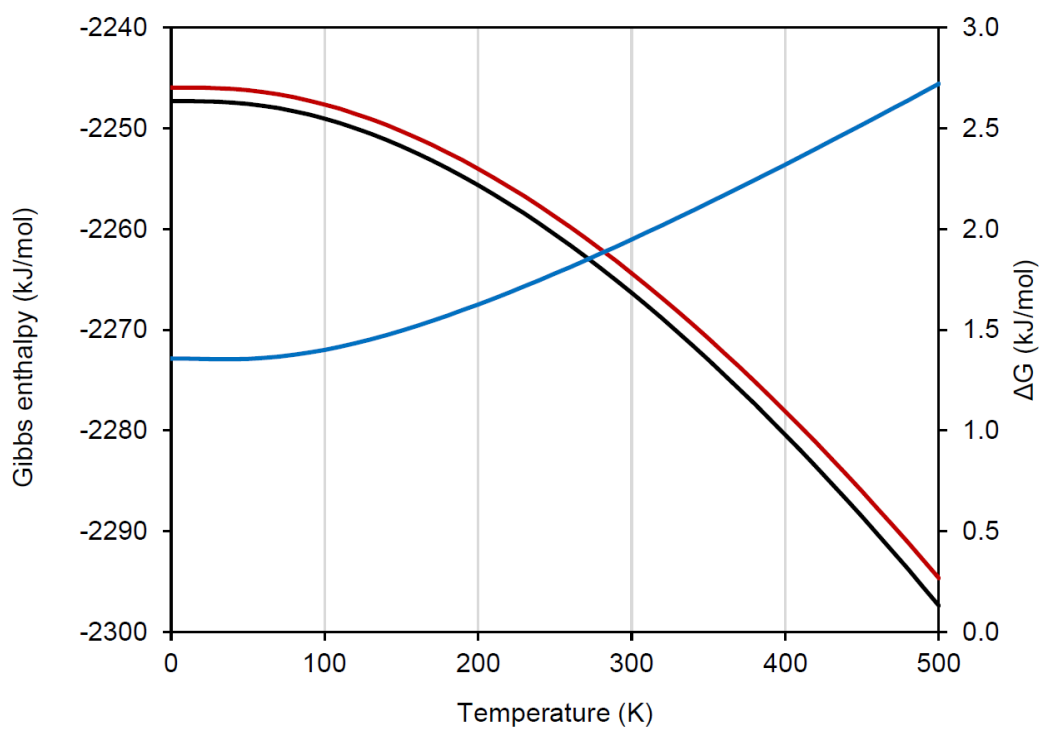

**Figure S9.** Gibbs enthalpy of α-AgSO$_4$ (red line) and β-AgSO$_4$ (black line) and their difference (ΔG = G$_α$ – G$_β$), calculated with the Phonopy[1] using VASP DFT+U approach as force calculator.

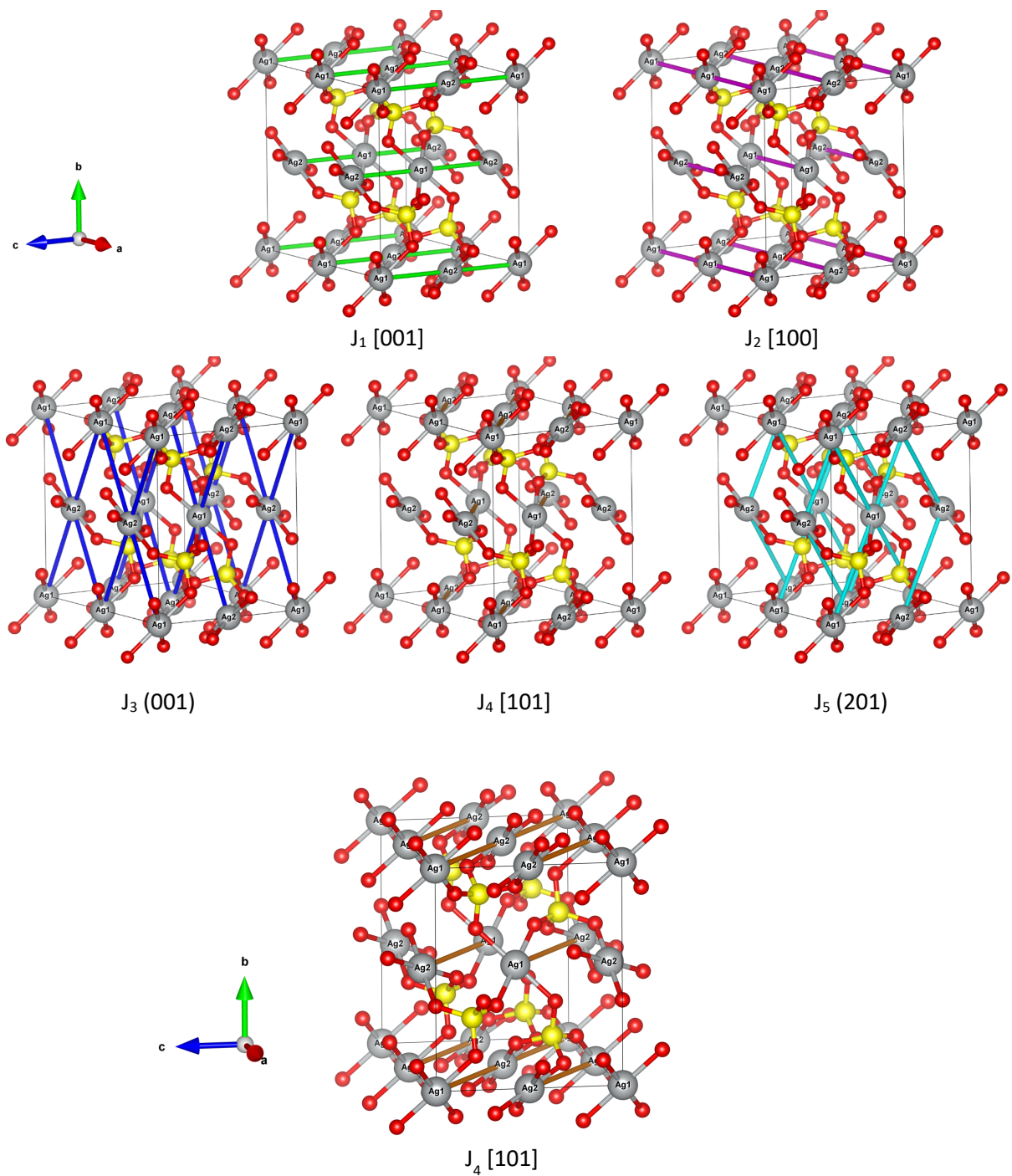

**Figure S10.** The view on the possible magnetic interactions in β-AgSO$_4$ crystal structure (supercell 2x1x1). Details are shown in the Tables S3-S5.

$J_3$ (001)

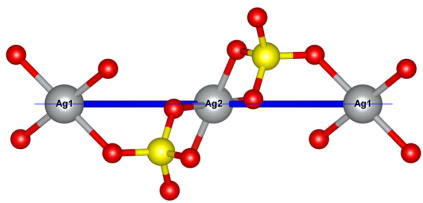 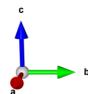 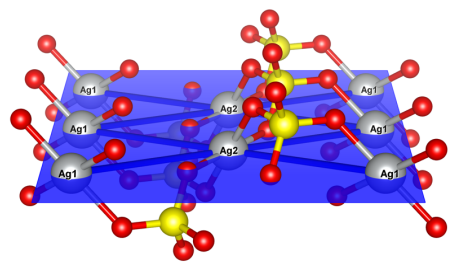

$J_{5\_1}$ (201)

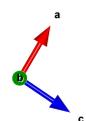 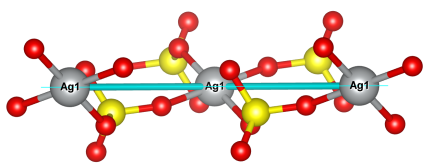 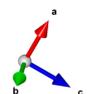 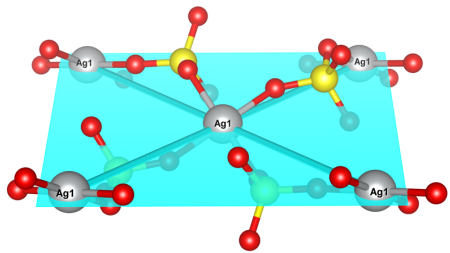

$J_{5\_2}$ (201)

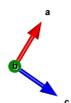 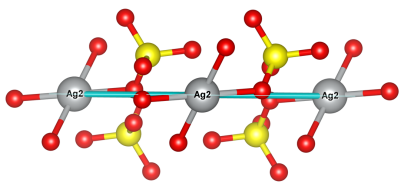 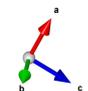 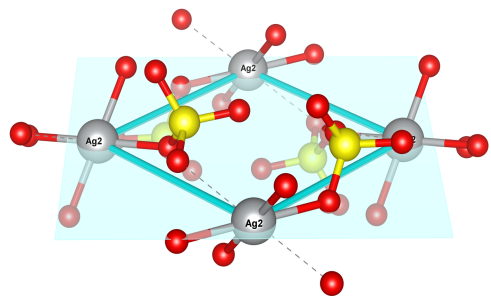

**Figure S11.** A close-up on the in-plane possible magnetic interactions in β-AgSO$_4$ crystal structure (supercell 2x1x1). Notice that $J_{5\_1}$ (between Ag1-Ag1) and $J_{5\_2}$ (between Ag2-Ag2) have different superexchange paths, so that these interactions can be separated.

**Table S3.** Interactions in the supercell 2x1x1 of (a) experimental structure of β-AgSO$_4$, (b) DFT+U calculated structure of β-AgSO$_4$ and (c) HSE06 calculated structure of β-AgSO$_4$. Notice we can distinguish between "mixed" interactions (i.e. Ag1-Ag2, grey background color), and "mixed" interactions (i.e. Ag1-Ag1, Ag2-Ag2, white background color). Asterisks indicate Ag-Ag contacts without at least one "strong" Ag-O bond (i.e. shorter than 2.3 Å). Column color relates to the interactions pictured in Figures S7-S8. "Prediction" stands for a qualitative prediction based on Goodenough-Kanamori rules.

a)

| Name | Type | Direction | Ag'-Ag" distance [Å] | Ag'-S-Ag" angle [deg.] | Prediction | Color |
|---|---|---|---|---|---|---|
| J$_1$ | Ag1-Ag2 | [001] | 3.601 | 65.7 | ferro weak | green |
| J$_{2\_1}$ | Ag1-Ag1 | [100] | 4.741 | (82.94)* | weak | magenta |
| J$_{2\_2}$ | Ag2-Ag2 | [100] | 4.741 | 97.3 | AF moderate | magenta |
| J$_3$ | Ag1-Ag2 | (001) | 4.976 | 103.0 | AF moderate | blue |
| J$_4$ | Ag1-Ag2 | [101] | 5.745 | 122.2 | AF strong | orange |
| J$_{5\_1}$ | Ag1-Ag1 | (201) | 6.042 | 132.4 | AF strong | cyan |
| J$_{5\_2}$ | Ag2-Ag2 | (201) | 6.042 | (128.4)* | weak | cyan |

b)

| Name | Type | Direction | Ag'-Ag" distance [Å] | Ag'-S-Ag" angle [deg.] | J / meV | Color |
|---|---|---|---|---|---|---|
| J$_1$ | Ag1-Ag2 | [001] | 3.675 | 66.5 | 6.7 | green |
| J$_{2\_1}$ | Ag1-Ag1 | [100] | 4.849 | (83.9)* | -2.5 | magenta |
| J$_{2\_2}$ | Ag2-Ag2 | [100] | 4.849 | 94.* | -27.6 | magenta |
| J$_3$ | Ag1-Ag2 | (001) | 5.087 | 102.2 | -3.3 | blue |
| J$_4$ | Ag1-Ag2 | [101] | 5.810 | 120.7 | -36.2 | orange |
| J$_{5\_1}$ | Ag1-Ag1 | (201) | 6.144 | 135.6 | -31.7 | cyan |
| J$_{5\_2}$ | Ag2-Ag2 | (201) | 6.144 | (127.1)* | 3.3 | cyan |

c)

| Name | Type | Direction | Ag'-Ag" distance [Å] | Ag'-S-Ag" angle [deg.] | J / meV | Color |
|---|---|---|---|---|---|---|
| J$_1$ | Ag1-Ag2 | [001] | 3.673 | 66.5 | 5.9 | green |
| J$_{2\_1}$ | Ag1-Ag1 | [100] | 4.841 | (83.8)* | 1.4 | magenta |
| J$_{2\_2}$ | Ag2-Ag2 | [100] | 4.841 | 95.0 | -20.4 | magenta |
| J$_3$ | Ag1-Ag2 | (001) | 5.083 | 102.3 | -4.6 | blue |
| J$_4$ | Ag1-Ag2 | [101] | 5.809 | 120.9 | -38.9 | orange |
| J$_{5\_1}$ | Ag1-Ag1 | (201) | 6.143 | 135.2 | -38.1 | cyan |
| J$_{5\_2}$ | Ag2-Ag2 | (201) | 6.143 | (127.3)* | 2.1 | cyan |

**Table S4.** Energy and silver spins for the 2x1x1 supercell of (a) DFT+U calculated structure of β-AgSO$_4$ and (b) HSE06 calculated structure of β-AgSO$_4$.

a)

| State | E (DFT+U) / eV | S$^{Ag1\_1}$ | S$^{Ag1\_2}$ | S$^{Ag1\_3}$ | S$^{Ag1\_4}$ | S$^{Ag2\_1}$ | S$^{Ag2\_2}$ | S$^{Ag2\_3}$ | S$^{Ag2\_4}$ |
|---|---|---|---|---|---|---|---|---|---|
| AFM1 | -190.21832621 | 0.5 | 0.5 | 0.5 | 0.5 | -0.5 | -0.5 | -0.5 | -0.5 |
| AFM2 | -190.21646496 | 0.5 | 0.5 | -0.5 | -0.5 | 0.5 | 0.5 | -0.5 | -0.5 |
| AFM3 | -190.26205085 | 0.5 | 0.5 | -0.5 | -0.5 | -0.5 | -0.5 | 0.5 | 0.5 |
| AFM4 | -190.28372692 | 0.5 | -0.5 | 0.5 | -0.5 | 0.5 | -0.5 | 0.5 | -0.5 |
| AFM5 | -190.19789881 | 0.5 | -0.5 | 0.5 | -0.5 | -0.5 | 0.5 | -0.5 | 0.5 |
| AFM6 | -190.27008567 | 0.5 | 0.5 | -0.5 | -0.5 | 0.5 | -0.5 | 0.5 | -0.5 |
| FM4_1 | -190.17578995 | 0.5 | 0.5 | 0.5 | 0.5 | 0.5 | 0.5 | -0.5 | -0.5 |
| FM4_2 | -190.24575830 | 0.5 | 0.5 | -0.5 | -0.5 | 0.5 | 0.5 | 0.5 | 0.5 |
| FM8 | -190.14099785 | 0.5 | 0.5 | 0.5 | 0.5 | 0.5 | -0.5 | 0.5 | -0.5 |

b)

| State | E (HSE06) / eV | S$^{Ag1\_1}$ | S$^{Ag1\_2}$ | S$^{Ag1\_3}$ | S$^{Ag1\_4}$ | S$^{Ag2\_1}$ | S$^{Ag2\_2}$ | S$^{Ag2\_3}$ | S$^{Ag2\_4}$ |
|---|---|---|---|---|---|---|---|---|---|
| AFM1 | -348.95328618 | 0.5 | 0.5 | 0.5 | 0.5 | -0.5 | -0.5 | -0.5 | -0.5 |
| AFM2 | -348.95923079 | 0.5 | 0.5 | -0.5 | -0.5 | 0.5 | 0.5 | -0.5 | -0.5 |
| AFM3 | -349.00683865 | 0.5 | 0.5 | -0.5 | -0.5 | -0.5 | -0.5 | 0.5 | 0.5 |
| AFM4 | -349.01080849 | 0.5 | -0.5 | 0.5 | -0.5 | 0.5 | -0.5 | 0.5 | -0.5 |
| AFM5 | -348.92125496 | 0.5 | -0.5 | 0.5 | -0.5 | -0.5 | 0.5 | -0.5 | 0.5 |
| AFM6 | -349.00559147 | 0.5 | 0.5 | -0.5 | -0.5 | 0.5 | -0.5 | 0.5 | -0.5 |
| FM4_1 | -348.90675906 | 0.5 | 0.5 | 0.5 | 0.5 | 0.5 | 0.5 | -0.5 | -0.5 |
| FM4_2 | -348.98729638 | 0.5 | 0.5 | -0.5 | -0.5 | 0.5 | 0.5 | 0.5 | 0.5 |
| FM8 | -348.86552869 | 0.5 | 0.5 | 0.5 | 0.5 | 0.5 | 0.5 | 0.5 | 0.5 |

**Table S5.** (a) Equations used in the Hamiltonian in the 2x1x1 supercell of β-AgSO$_4$ structure together with (b) their solutions for each of the magnetic states presented in Table S4. The determinants method was used to solve linear equations systems for each case, the results are presented in Table S3.

a)

$J_1 = -(S^{Ag1\_1} \cdot S^{Ag2\_1} + S^{Ag1\_2} \cdot S^{Ag2\_2} + S^{Ag1\_3} \cdot S^{Ag2\_3} + S^{Ag1\_4} \cdot S^{Ag2\_4})$

$J_{2\_1} = -(S^{Ag1\_1} \cdot S^{Ag1\_2} + S^{Ag1\_3} \cdot S^{Ag1\_4})$

$J_{2\_2} = -(S^{Ag2\_1} \cdot S^{Ag2\_2} + S^{Ag2\_3} \cdot S^{Ag2\_4})$

$J_3 = -((S^{Ag1\_1} + S^{Ag1\_2}) \cdot (S^{Ag2\_3} + S^{Ag2\_4}) + (S^{Ag1\_3} + S^{Ag1\_4}) \cdot (S^{Ag2\_1} + S^{Ag2\_2}))$

$J_4 = -(S^{Ag1\_1} \cdot S^{Ag2\_2} + S^{Ag1\_2} \cdot S^{Ag2\_1} + S^{Ag1\_3} \cdot S^{Ag2\_4} + S^{Ag1\_4} \cdot S^{Ag2\_3})$

$J_{5\_1} = -((S^{Ag1\_1} + S^{Ag1\_2}) \cdot (S^{Ag1\_3} + S^{Ag1\_4}))$

$J_{5\_2} = -((S^{Ag2\_1} + S^{Ag2\_2}) \cdot (S^{Ag2\_3} + S^{Ag2\_4}))$

b)

AFM1 = 2 E$_c$ + 1 J$_1$ − 0.5 J$_{2\_1}$ − 0.5 J$_{2\_2}$ + 2 J$_3$ + 1 J$_4$ − 1 J$_{5\_1}$ − 1 J$_{5\_2}$

AFM2 = 2 E$_c$ − 1 J$_1$ − 0.5 J$_{2\_1}$ − 0.5 J$_{2\_2}$ + 2 J$_3$ − 1 J$_4$ + 1 J$_{5\_1}$ + 1 J$_{5\_2}$

AFM3 = 2 E$_c$ + 1 J$_1$ − 0.5 J$_{2\_1}$ − 0.5 J$_{2\_2}$ − 2 J$_3$ + 1 J$_4$ + 1 J$_{5\_1}$ + 1 J$_{5\_2}$

AFM4 = 2 E$_c$ − 1 J$_1$ + 0.5 J$_{2\_1}$ + 0.5 J$_{2\_2}$ + 0 J$_3$ + 1 J$_4$ + 0 J$_{5\_1}$ + 0 J$_{5\_2}$

AFM5 = 2 E$_c$ + 1 J$_1$ + 0.5 J$_{2\_1}$ + 0.5 J$_{2\_2}$ + 0 J$_3$ − 1 J$_4$ + 0 J$_{5\_1}$ + 0 J$_{5\_2}$

AFM6 = 2 E$_c$ + 0 J$_1$ − 0.5 J$_{2\_1}$ + 0.5 J$_{2\_2}$ + 0 J$_3$ + 0 J$_4$ + 1 J$_{5\_1}$ + 0 J$_{5\_2}$

FM4_1 = 2 E$_c$ + 0 J$_1$ − 0.5 J$_{2\_1}$ − 0.5 J$_{2\_2}$ + 0 J$_3$ + 0 J$_4$ − 1 J$_{5\_1}$ + 1 J$_{5\_2}$

FM4_2 = 2 E$_c$ + 0 J$_1$ − 0.5 J$_{2\_1}$ − 0.5 J$_{2\_2}$ + 0 J$_3$ + 0 J$_4$ + 1 J$_{5\_1}$ − 1 J$_{5\_2}$

FM8 = 2 E$_c$ − 1 J$_1$ − 0.5 J$_{2\_1}$ − 0.5 J$_{2\_2}$ − 2 J$_3$ − 1 J$_4$ − 1 J$_{5\_1}$ − 1 J$_{5\_2}$

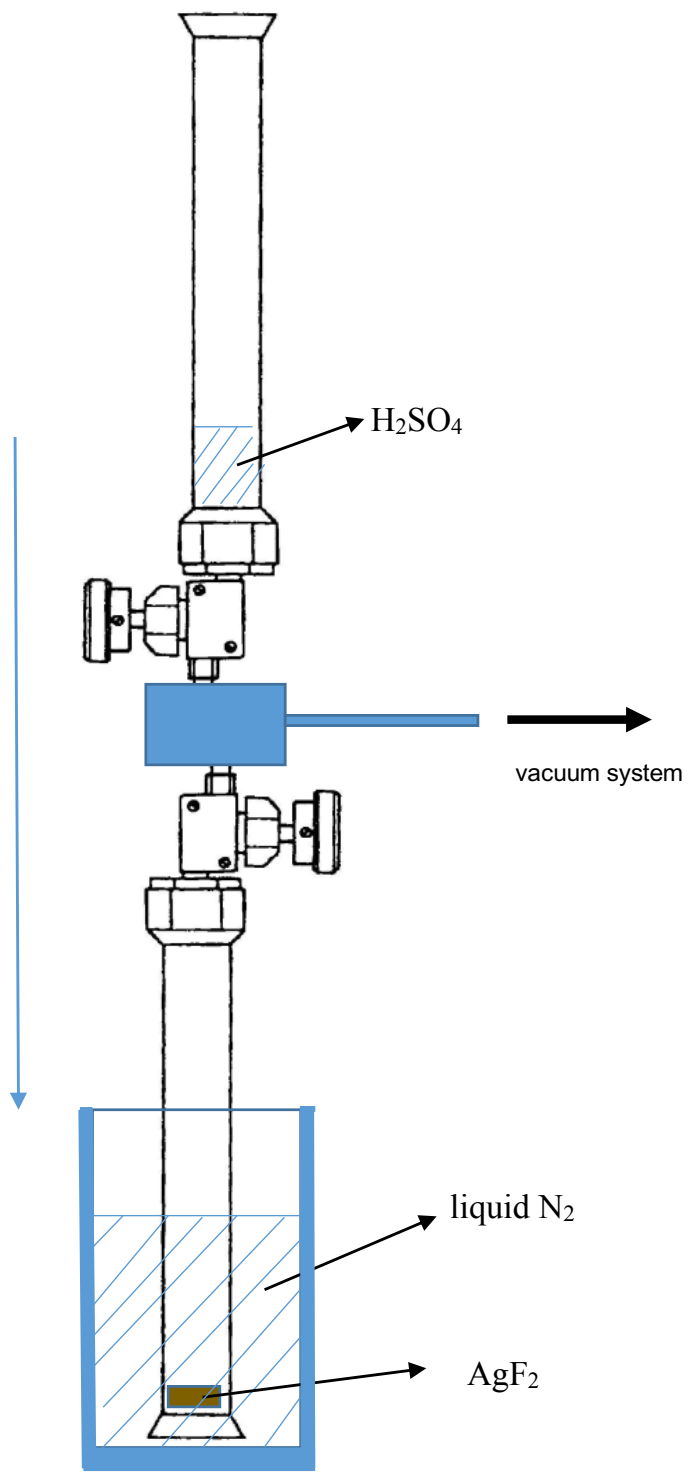

**Figure S12.** Experimental setup for the reaction between AgF₂ and H₂SO₄.

**SI References**